\documentclass[a4, useAMS,usenatbib]{mn2e}

\pdfoutput=1
\usepackage[pdftex]{graphicx}
\usepackage[pdftex]{color}
\usepackage[colorlinks,bookmarks]{hyperref}
\definecolor{linkblue}{rgb}{0,0,0.8}
\definecolor{linkgreen}{rgb}{0,0.5,0}
\hypersetup{linkcolor=linkblue, citecolor=linkgreen, urlcolor=linkblue}

\usepackage{textcase}
\usepackage[T1]{fontenc}
\usepackage[utf8]{inputenc}
\usepackage{lmodern}
\usepackage[english]{babel}
\usepackage{bm}
\usepackage{verbatim}
\usepackage[us]{datetime}
\usepackage{amsmath, amssymb}
\usepackage{nicefrac}
\usepackage{enumerate}

\providecommand{\eprint}[1]{\href{http://arxiv.org/abs/#1}{#1}}
\providecommand{\adsurl}[1]{\href{#1}{ADS}}

\usepackage{ifthen}\def\eprinttmp@#1arXiv:#2 [#3]#4@{
\ifthenelse{\equal{#3}{x}}{\href{http://arxiv.org/abs/#1}{#1}
}{\href{http://arxiv.org/abs/#2}{arXiv:#2} [#3]}}
\renewcommand{\eprint}[1]{\eprinttmp@#1arXiv: [x]@}

\def\aap{A\&A}

\def\apjl{ApJ}

\def\apjs{ApJS}

\newcommand{\omegam}{\Omega_{m0}}
\newcommand{\omegamz}{\Omega_{m}(z)}
\newcommand{\sig}{\sigma_8}
\newcommand{\siga}{\sigma_{8,\gamma}}

\newcommand{\gas}{\gamma_{\rm sm}}

\newcommand{\geds}{G_{\scriptscriptstyle \rm EdS}}

\newcommand{\dd}{\textrm{d}}

\newcommand{\tgl}{\texttt{turboGL}\ }

\newcommand{\threeint}{\mu_{3,\rm int}}
\newcommand{\fourint}{\mu_{4,\rm int}}

\title[Constraining perturbations with supernovae]{Constraining the growth of perturbations with lensing of supernovae}

\author[Amendola, Castro, Marra and Quartin]{Luca Amendola$^{1}$, Tiago Castro$^{2}$, Valerio Marra$^{2,3}$ and Miguel Quartin$^{2}$\\
$^{1}$Institut für Theoretische Physik, Universität Heidelberg, Philosophenweg 16, 69120 Heidelberg, Germany\\
$^{2}$Instituto de Física, Universidade Federal do Rio de Janeiro, CEP 21941-972, Rio de Janeiro, RJ, Brazil\\
$^{3}$Departamento de Física, Universidade Federal do Esp\'{\i}rito Santo, %Av.~F.~Ferrari, 514,
29075-910, Vit\'oria, ES, Brazil}

\begin{document}
%%%%%%%%%%%%%%%%%%%%%%%%%%%%%%%%%%%%
%%%%%%%%%%%%%%%%%%%%%%%%%%%%%%%%%%%%

\date{Accepted XXX. Received XXX; in original form 17/Feb/2015}

\pagerange{\pageref{firstpage}--\pageref{lastpage}} \pubyear{2015}

\maketitle

\label{firstpage}

\begin{abstract}
A recently proposed technique allows one to constrain both the background and perturbation cosmological parameters through the distribution function of supernova Ia apparent magnitudes.
Here we extend this technique to alternative cosmological scenarios, in which the growth of structure does not follow the $\Lambda$CDM prescription.
We apply the method first to the supernova data provided by the JLA catalog combined with all the current independent redshift distortion data and with low-redshift cluster data from Chandra and show that although the supernovae alone are not very constraining, they help in reducing the confidence regions. Then we apply our method to future data from LSST and from a survey that approximates the Euclid satellite mission. In this case we show that the combined data are nicely complementary and can constrain the normalization $\sigma_8$ and the growth rate index $\gamma$ to within $0.6\%$ and $7\%$, respectively. In particular, the LSST supernova catalog is forecast to give the constraint $\gamma (\sigma_8/0.83)^{6.7} = 0.55 \pm 0.1$. We also report on constraints relative to a step-wise parametrization of the growth rate of structures. These results show that supernova lensing serves as a good cross-check on the measurement of perturbation parameters from more standard techniques.
\end{abstract}


%%%%%%%%%%%%%%%%%%%%%%%%%%%%%%%%%%%%
%%%%%%%%%%%%%%%%%%%%%%%%%%%%%%%%%%%%
\section{Introduction}
\label{intro}
%%%%%%%%%%%%%%%%%%%%%%%%%%%%%%%%%%%%
%%%%%%%%%%%%%%%%%%%%%%%%%%%%%%%%%%%%

The need to understand the nature of the mechanism that accelerates the Universe expansion is driving
several new observational campaigns. Dedicated surveys like the %DES
Dark Energy Survey
or satellites like Euclid, along with several
other ground-based  surveys, will soon generate very large databases of galaxy redshifts and images and of
supernova lightcurves, all of them containing information of cosmological dynamics.

In this paper,  following a series of previous works \citep{Amendola:2013twa, Quartin:2013moa, Castro:2014oja}, we exploit a complementary method based on lensing of distant supernovae Ia (SNe Ia). The idea behind our approach is rather simple. The apparent magnitude of standard candles
depends on both the background cosmology and the gravitational perturbations crossed by the photons along their path; by analyzing the statistical distribution of the SN Ia apparent magnitudes around their mean value we can infer the properties of the intervening  matter perturbations. This follows ideas first discussed in~\citealt{Bernardeau:1996un,Hamana:1999rk,Valageas:1999ir} and later further developed in~\citealt{Dodelson:2005zt}.
In practice some assumptions on the intrinsic scatter of SNe Ia are required; in particular, we assume that the intrinsic scatter is independent of redshift, since this was shown to be the most reasonable hypothesis using Bayesian analysis \citep{Castro:2014oja} and also agrees with the underlying motivation for using SNe Ia as standard candles whose properties do not depend on distance.

As mentioned earlier, supernova lensing depends both on background and  perturbation parameters. Building $N$-body simulations for a reasonable grid of cosmological parameters and then extracting the theoretical  magnitude distribution is hardly feasible. To circumvent this limitation we employed realizations of the perturbed universe using the \texttt{turboGL}\footnote{The work carried out in this paper is based on (the latest) version 3.0 of \tgl available at \href{http://www.turbogl.org/}{turbogl.org}} implementation of sGL (stochastic Gravitational Lensing) -- a very fast method developed by~\citealt{Kainulainen:2009dw,Kainulainen:2010at,Kainulainen:2011zx} -- whose results are in concordance with recent $N$-body simulations but orders of magnitude faster \citep{Amendola:2013twa}. They are also in very good agreement with observational data \citep{Jonsson:2009jp,Kronborg:2010uj,Jonsson:2010wx} and with other recent independent theoretical estimations \citep{BenDayan:2013gc}.

In our previous papers we discussed how the SN Ia magnitude scatter can be employed to constrain cosmological parameters within the standard model of cosmology. In a $\Lambda$CDM scenario it was confirmed that the matter density $\Omega_{m0}$ and the amplitude of the power spectrum $\sigma_8$ were the most important cosmological parameters as far as supernova lensing is concerned.
As the former is tightly constrained by the supernova magnitudes themselves (i.e., by the first moment of the distribution), the most important \emph{new} information gained was the value of $\sigma_8$. In this respect we found that while present catalogs start to have the statistical power to make the first measurement of $\sig$ \citep{Castro:2014oja}, LSST will be able to constrain the amplitude of the power spectrum at the level of a few percents with supernovae only \citep{Quartin:2013moa}.

Here we extend our previous analysis to the case of a non-standard growth rate, which we model using either the $f_g \approx \omegamz^\gamma$ or the step-wise $f_g = f_i$ parametrization. We then apply the Method-of-Moments (MeMo, see \citealt{Quartin:2013moa}) -- which basically compares the observed central moments of the magnitude distribution to the theoretical predictions -- to current and future supernova catalogs.
Regarding the latter we base our study on the Large Synoptic Survey Telescope (LSST, see \citealt{Abell:2009aa}) and the Wide-Field Infrared Survey Telescope (WFIRST, see \citealt{Green:2012mj}). Regarding current data, we use the most recent SN catalog dubbed JLA (acronym for Joint Lightcurve Analysis, see \citealt{Betoule:2014frx}).

In order to illustrate the complementarity of our method we combine the current results on $\sig$ and $\gamma$ with low-redshift cluster data from Chandra (see \citealt{Vikhlinin:2008ym}) and with redshift distortion data from all the current independent surveys: 2dFGS, 6dFGS, LRG, BOSS, CMASS, WiggleZ and
VIPERS (see respectively, \citealt{Percival:2004fs}, \citealt{Beutler:2012px}, \citealt{Samushia:2011cs,Chuang:2012qt}, \citealt{Tojeiro:2012rp},  \citealt{Beutler:2013yhm,Reid:2014iaa,Samushia:2013yga}, \citealt{Blake:2012pj} and \citealt{delaTorre:2013rpa}); and our forecast results with constraints from a survey that approximates the Euclid satellite mission (see, \citealt{Euclid-r,Amendola:2012ys}).

It is clear that the large increase in supernova statistics provided by the LSST should be accompanied by a similar increase on our understanding of the various SN systematics. In fact in the past few years a large effort has been devoted to testing and improving the calibration of SN Ia and to correcting their light curves in order to understand and control systematics \citep{Kessler:2009ys,Conley:2011ku,Betoule:2012an,Scolnic:2013aya}. In the forecasts here presented we assume that systematics can be kept subdominant even for LSST, but it is not at all clear if this will be achieved.

This paper is organized as follows. In Section~\ref{model} we will state the adopted model for background and perturbations, and also the method we use to compute the lensing distribution. In Section~\ref{meda} we will build the likelihood functions for the various observables and the corresponding data, while in Section~\ref{constraints} we will show the constraining power of supernova catalogs. Conclusions are drawn in Section~\ref{conclusions}. Finally, in Appendix~\ref{app:fits} simple fits for the lensing moments as a function of $\{z,\,\sigma_{8},\,\gamma\}$ will be given.

%%%%%%%%%%%%%%%%%%%%%%%%%%%%%%%%%%%%
%%%%%%%%%%%%%%%%%%%%%%%%%%%%%%%%%%%%
\section{Model} \label{model}
%%%%%%%%%%%%%%%%%%%%%%%%%%%%%%%%%%%%
%%%%%%%%%%%%%%%%%%%%%%%%%%%%%%%%%%%%

%%%%%%%%%%%%%%%%%%%%%%%%%%%%%%%%%%%%
\subsection{Matter and lensing model}
\label{lensing}
%%%%%%%%%%%%%%%%%%%%%%%%%%%%%%%%%%%%

We will obtain the lensing probability density function (PDF) for the desired model parameters using the \tgl code,
which is the numerical implementation of the stochastic gravitational lensing (sGL) method introduced in \citealt{Kainulainen:2009dw,Kainulainen:2010at,Kainulainen:2011zx}.
The sGL method is based on (i) the weak lensing approximation and (ii) generating stochastic configurations of inhomogeneities along the line of sight.

Regarding (ii), the matter density contrast $\delta_{M}(r,t)$ is modeled in the present paper as a random collection of Navarro-Frenk-White (NFW) halos, whose abundance is calculated using the halo mass function given in \citealt{Courtin:2010gx} (basically a refitted \citealt{Sheth:1999mn} mass function) and whose concentration parameters (which depend on cosmology) are calculated using the universal and accurate model proposed in \citealt{Zhao:2008wd}.
Linear correlations in the halo positions are neglected by the sGL method: this should be indeed a good approximation as the contribution of the 2-halo term is negligible with respect to the contribution of the 1-halo term \citep{Kainulainen:2011zx}.
Overall, the modeling was proved \citep{Amendola:2013twa} to be accurate for the redshift range of $z \lesssim 1.5$ in which we are mainly interested in this paper. At higher redshifts the relative importance of unvirialized objects such as filaments becomes more important and one may need to include them in the modeling \citep{Kainulainen:2010at}.

Regarding (i), the lens convergence $\kappa$ in the weak-lensing approximation is given by the following line-of-sight integral \citep{Bartelmann:1999yn}:
\begin{equation} \label{eq:kappa}
    \kappa(z_{s})=\rho_{MC} \, \int_{0}^{r_{s}}dr \, {\cal G}(r,r_{s})\,\delta_{M}(r,t(r)) \,,
\end{equation}
where the quantity $\delta_{M}(r,t)$ is the local matter density contrast (which is modeled as described above), $\rho_{MC} \equiv a_0^3 \, \rho_{M0}$ is the constant matter density in a comoving volume, and the function $${\cal G}(r,r_{s})=  \frac{4\pi G}{c^2 \, a}  \; \frac{r (r_{s}-r)}{r_{s}}$$ gives the optical weight of a matter structure at the comoving radius $r$ (assuming spatial flatness).
The functions $a(t)$ and $t(r)$ are the scale factor and geodesic time for the background FLRW model, and $r_{s}=r(z_{s})$ is the comoving position of the source at redshift $z_{s}$.
At the linear level, the shift in the distance modulus caused by lensing is expressed in terms of the convergence only:
\begin{equation} \label{eq:dm}
\Delta m(z) \simeq - \frac{5}{\log 10} \; \kappa(z) \,.
\end{equation}
Eq.~(\ref{eq:kappa}) connects the statistics of the matter distribution to the
statistics of the convergence distribution: by studying the latter one can gain information on the former and thus on the nature of dark energy.
The sGL method for computing the lens convergence is based on generating random
configurations of halos along the line of sight and computing the associated
integral in Eq.~(\ref{eq:kappa}) by binning into a number of independent lens
planes. A detailed explanation of the sGL method can be found
in \citealt{Kainulainen:2009dw,Kainulainen:2010at,Kainulainen:2011zx}.

Because of  theoretical approximations and modeling uncertainties, the \tgl code can be relied upon at the level of $\sim$10\% as far as the moments of the lensing PDF are concerned \citep{Amendola:2013twa}.

%%%%%%%%%%%%%%%%%%%%%%%%%%%%%%%%%%%%
\subsection{Growth of perturbations}
\label{grope}
%%%%%%%%%%%%%%%%%%%%%%%%%%%%%%%%%%%%

In this paper we aim at testing with supernova data the growth of perturbations. We will take a minimal and simple approach: we will assume that the background evolves according to the standard $\Lambda$CDM model but that matter perturbations grow according to a different theory. This is actually the case for most of the viable $f(R)$ and scalar-field models \citep{2010deto.book.....A}.
Therefore, while the supernova Hubble diagram in each redshift bin will have its \emph{mean} unchanged, it will feature a different \emph{dispersion} due to a different lensing caused by a different growth of structures.

In this Section we will discuss two ways to parametrize the growth rate of perturbations. Before starting, however, we would like to point out that we will use a halo mass function and concentration parameter model (see Section~\ref{lensing}) which have been tested within the standard paradigm. Therefore, our results may suffer from systematic errors when inspected far from the fiducial model, which we take to be the Planck 2013 best fit to observations of the cosmic microwave background (CMB) and baryon acoustic oscillations (BAO) \citep[][Table 5]{Ade:2013zuv}.

%%%%%%%%%%%%%%%%%%%%%%%%%%%%%%%%%%%%
\subsubsection{$\gamma$-parametrization}
%%%%%%%%%%%%%%%%%%%%%%%%%%%%%%%%%%%%

The linear growth of matter perturbations is described by the growth function $G(z)$, usually normalized to unity at the present time, $G(0)=1$. It is useful to describe the growth of perturbations via the growth rate $f_g$, which is the logarithmic derivative of the growth function with respect to the scale factor $a=1/(1+z)$:
\begin{equation}
    f_g \equiv - \frac{\dd \ln G}{\dd \ln (1+z)}  \approx \omegamz^\gamma \,.
    \label{gammapar}
\end{equation}
The last equation approximates the growth rate as a power of $\omegamz =\omegam (1+z)^3/ E^2(z)$, where $E(z)=H(z)/H_0$, $H$ is the Hubble parameter and the subscript ``0'' denotes the present-day value of a quantity.
Within General Relativity and for the $\Lambda$CDM model $f_g$ is accurately described by Eq.~(\ref{gammapar}) with $\gamma=\gas \approx 0.55$ \citep{2010deto.book.....A}, and the subject of Section~\ref{constraints} will be to understand how strongly can lensing of supernovae constrain $\gamma$ around this standard value. Indeed, any measured deviation from $\gas$ will signal the demise of the standard model of cosmology.

The growth function is obtained by the following integral of the growth rate:
\begin{equation}
    G(z) \,=\, \exp \left( -\int_0^z \frac{\dd \bar z}{1+ \bar z} f_g(\bar z) \right) .
    \label{g-function}
\end{equation}
In the Einstein-de Sitter (EdS) model -- which is effectively the same as $\Lambda$CDM at $1 \ll z \ll 1000$ -- one has $f_g=1$ and $\geds=1/(1+z)$.

%%%%%%%%%%%%%%%%%%%%%%%%%%%%%%%%%%%%
\subsubsection{$f$-parametrization}
\label{fpar}
%%%%%%%%%%%%%%%%%%%%%%%%%%%%%%%%%%%%

The parametrization of Eq.~(\ref{gammapar}), while certainly convenient, has a constrained redshift evolution and one may wish for a more general non-parametric description of the growth rate at the various redshifts. One possibility is to model the growth rate as a step-wise function \citep{Amendola:2012ys}:%
\begin{equation}
f_g=  f_i \,,
\label{fpareq}
\end{equation}
where $f_i$ is the value of the growth rate in the redshift bin $[z_{i-1}, z_i)$, where $z_0=0$ and $i=1, \dots, n$.
We will use bins of width $\Delta z=0.2$, which allows for a reconstruction -- with negligible error -- of the standard model growth function $G(z)$. The fiducial values of $f_i$ are given in Table~\ref{fbins}.

\begin{table}
\begin{center}
    \begin{tabular}{lccccc}
    \hline
    \hline
    $[z_{i-1}, z_i)$ & $\![0, 0.2)\!$ & $\![0.2, 0.4)\!$ & $\![0.4, 0.6)\!$ & $\![0.6, 0.8)\!$ & \!$[0.8, 1.0)\!$ \\
    $f_i$ & 0.58  & 0.68 & 0.76 & 0.81 & 0.86 \\
    \hline
    \hline
    \end{tabular}
\end{center}
\vspace{-.2cm}
\caption{Fiducial values of $f_{1,.., 5}$ of Eq.~(\ref{fpareq}), corresponding to $\omegamz^{\gas}$ calculated at the center of each redshift bin. \vspace{.1cm}\label{fbins}}
\end{table}%

The growth function, obtained performing the integral of Eq.~(\ref{g-function}), is then:
\begin{equation}
    G(z)= \left(\frac{1+z}{1+z_{\bar n}}\right)^{-f_{\bar n+1}}  \prod_{i=1}^{\bar n}\left(\frac{1+z_{i}}{1+z_{i-1}}\right)^{-f_{i}} ,
\end{equation}
where $\bar n$ is such that $z_{\bar n}< z$.

In Figure~\ref{fig:fG} an example of a step-wise growth rate and corresponding growth function are shown. Note that the nonstandard $G$ is higher in the past for an $f_g$ which is lower than in the $\Lambda$CDM in a low-z bin because $G$ is normalized to unity at present time.

It is worth stressing that the main feature of this parametrization is its flexibility. It allows our discussion to be very broad, and if one intends to test any modified gravity theory with the analysis developed here one needs only to derive the values of the $\{f_i\}$ set for the chosen theory and update the fiducial Table~\ref{fbins} .

%%%%%%%%%%%%%%%%%%%%%%%%%%%%%%%%%%%%
%%%%%%%%%%%%%%%%%%%%%%%%%%%%%%%%%%%%
\section{Method and data}
\label{meda}

\subsection{Supernova data}

\subsubsection{Method of the Moments}
\label{semo}
%%%%%%%%%%%%%%%%%%%%%%%%%%%%%%%%%%%%

\begin{figure}
\begin{centering}
    \includegraphics[width=0.95\columnwidth]{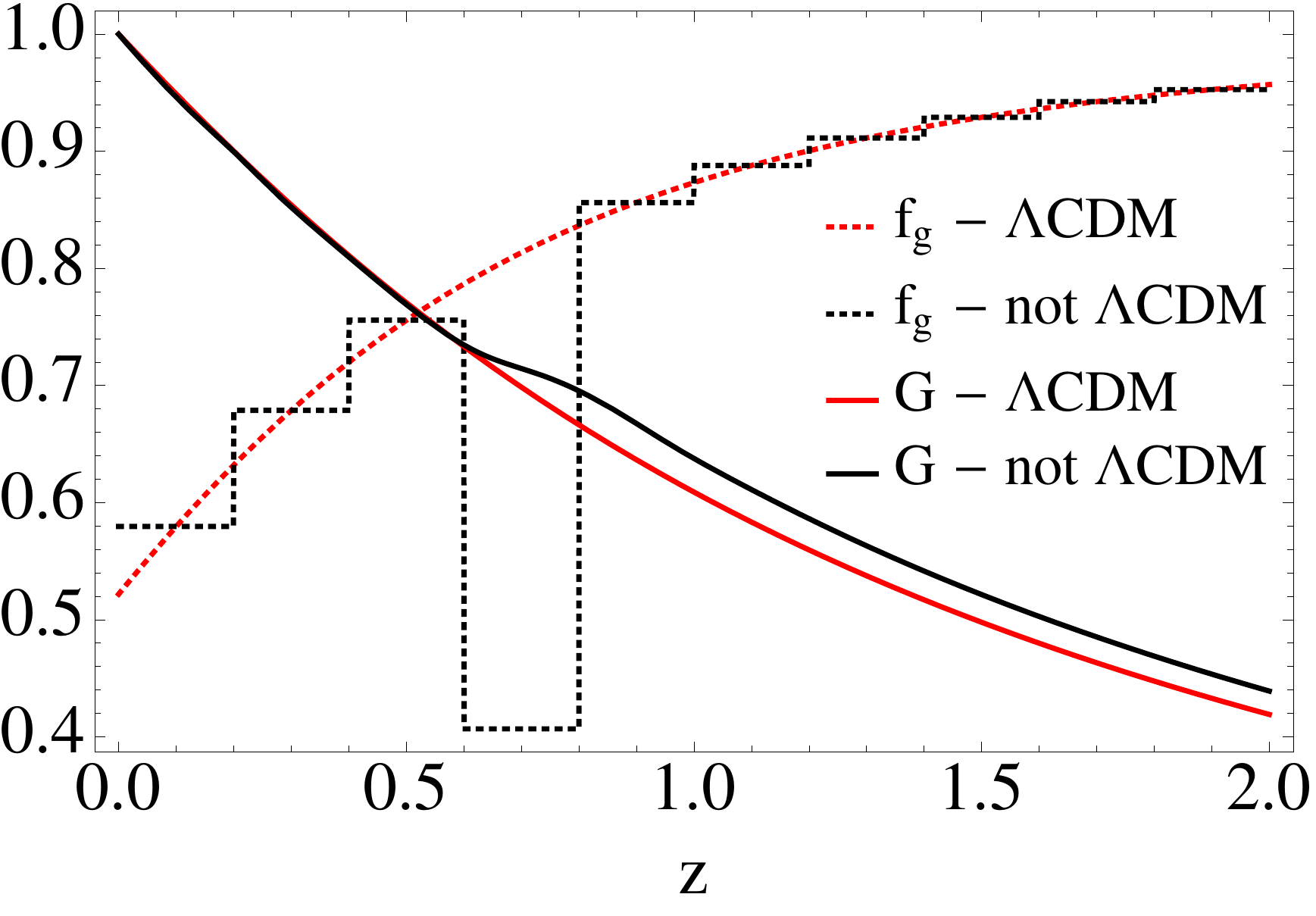}
    \caption{Growth function $G(z)$ and growth rate $f_g(z)$ for the $\Lambda$CDM model (red) and a model whose growth rate is parametrized according to a step-wise function (black) which has the standard value for every redshift bin except the bin $[0.6, 0.8)$ where $f_g$ has half its standard value. See Section~\ref{fpar} for more details.}
    \label{fig:fG}
\end{centering}
\end{figure}

Here we summarize the so-called \emph{Method-of-the-Moments} (MeMo). Originally discussed in \citealt{Quartin:2013moa} in the context of forecasts, the MeMo has recently passed its first test with real data as shown in \citealt{Castro:2014oja}. In a nutshell, the idea is to use the scatter in the Hubble diagram to measure cosmological parameters on which gravitational lensing depends: $\omegam$ and $\sig$ in \citealt{Quartin:2013moa,Castro:2014oja} and also  $\gamma$ or $f_i$ in the present work. The MeMo approach is basically a $\chi^2$ approach where we measure the mean $\mu_{1}'$ (which is independent of lensing due to photon number conservation) and the first three central moments $\{\mu_{2},\mu_{3},\mu_{4}\}$ (which we will collectively refer to simply as $\mu_{1-4}$) and compare them with the corresponding theoretical predictions.
The latter is computed through the convolution of the lensing PDF ($\mu_{1-4,\rm lens}$, see Section~\ref{lensing}) with the intrinsic SN dispersion distribution ($\{\sigma_{\rm int},\threeint,\fourint\}$, which we define including all instrumental noise contributions). The final relation between those quantities are given by \citep{Quartin:2013moa}:
\begin{align}
    \mu_{2} & \;\equiv\;\sigma_{{\rm tot}}^{2}
    \;=\;\sigma_{{\rm lens}}^{2}+\sigma_{\rm int}^{2}\,,\label{eq:mu2}\\
    \mu_{3} & \;=\;\mu_{3,{\rm lens}} + \threeint \,,\label{eq:mu3}\\
    \mu_{4} & \;=\;\mu_{4,{\rm lens}}+6\,\sigma_{{\rm lens}}^{2}\,\sigma_{\rm int}^{2} +  3 \, \sigma_{\rm int}^{4} + \fourint \,.\label{eq:mu4}
\end{align}
Although the number of moments to be used in the analysis is in principle arbitrary as each new moment adds more information, it was shown in \citealt{Quartin:2013moa} that for supernova analyses  basically all the information is already contained in the first four moments $\mu_{1-4}$ (and a very good fraction of it already in $\mu_{1-3}$).

The full MeMo likelihood is then:
\begin{align}
    &L_{\rm MeMo}({\rm Data}|\Theta_{\rm{cosmo}}) = \exp \bigg( - \frac{1}{2} \sum_{j}^{{\rm bins}} \chi_{j}^2 \bigg) \,,\label{Lmom} \\
    &\chi^2_j = \big(\boldsymbol{\mu}-\boldsymbol{\mu}_{\rm data}\big)^t \;\Sigma_j^{-1}\; \big(\boldsymbol{\mu}-\boldsymbol{\mu}_{\rm data}\big) \,, \label{chi2mom} \\
    &\boldsymbol{\mu} = \{ \mu_1',\,\mu_2,\,\mu_3,\,\mu_4 \} \,,
\end{align}
where the vector $\boldsymbol{\mu}$ depends on the cosmological parameters $(\Theta_{\rm{cosmo}})$, and its second-to-fourth components are defined in~Eqs~\eqref{eq:mu2}-\eqref{eq:mu4}. The mean $\mu_{1}'$ is the  theoretical distance modulus.
The components of the vector $\boldsymbol{\mu}_{\rm data}(z_{j})$ are the moments inferred from the data, which for the forecasts we take to be $\boldsymbol{\mu}(\Theta_{\rm{cosmo}})$ evaluated at the fiducial model and at redshift $z_{j}$. The covariance matrix $\Sigma$ is also built using the fiducial moments  and therefore does not depend explicitly on cosmology (but it does on $z$ --  see \citealt{Quartin:2013moa} for more details).

Even though the (most recent) JLA supernova catalog \citep{Betoule:2014frx} showed no need of intrinsic moments higher than the second \citep{Castro:2014oja}, with more accurate and precise surveys new systematic effects may become evident. Therefore, in order to obtain conservative results we allow the intrinsic supernova distribution to also have non-zero intrinsic third ($\threeint$) and fourth ($\fourint$) moment. In the case of the JLA catalog the statistics are not good enough to warrant $\fourint$ \citep{Castro:2014oja}, but we keep it in the forecast analysis. Also, as in \citep{Castro:2014oja} here we assume an intrinsic distribution constant in redshift since this was shown to be currently the most reasonable hypothesis using Bayesian analysis.

%%%%%%%%%%%%%%%%%%%%%%%%%%%%%%%%%%%%
\subsubsection{Supernova catalogs}
\label{cats}
%%%%%%%%%%%%%%%%%%%%%%%%%%%%%%%%%%%%

In this section we describe the main details of the real and synthetic catalogs which we will use throughout this paper. We base our study on two future surveys -- the Large Synoptic Survey Telescope (LSST, see \citealt{Abell:2009aa}) and the Wide-Field Infrared Survey Telescope (WFIRST, see \citealt{Green:2012mj}) -- and the most recent SN catalog dubbed JLA (acronym for Joint Lightcurve Analysis, see \citealt{Betoule:2014frx}).

\begin{figure}
\begin{centering}
\includegraphics[width=\columnwidth]{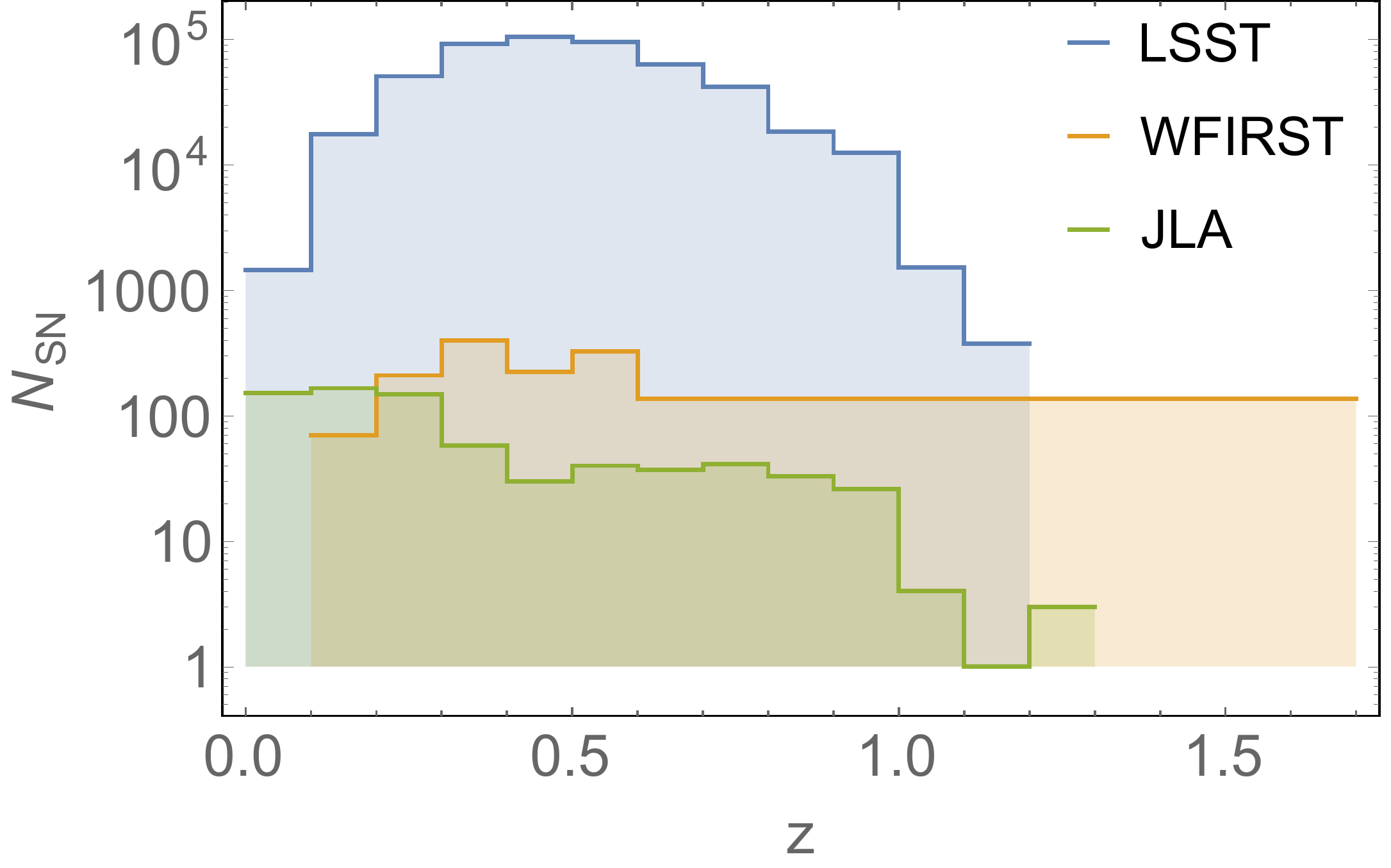}
    \caption{Observed supernova redshift distributions of the JLA dataset of 740 SNe and forecast distributions for LSST (5 years of observations for a total of 500,000 SNe, although this depend on the imposed quality cuts) and WFIRST (2700 SNe). See Section~\ref{cats} for more details.}
\label{fig:SNz}
\end{centering}
\end{figure}

LSST is an upcoming photometric survey currently in the design and development phase and expected to be operational between the end of this decade and the beginning of the next. By the end of its ten-year mission the number of supernovae observed will be a few millions. This number includes all the expected observed supernovae but in this paper we adopt the distribution based on the selection cut of signal to noise higher than fifteen in at least two filters. With that cut the total number of supernovae decreases to half a million in five years, and this was the number used here when computing the SN distribution shown in Figure~\ref{fig:SNz} (we include SN from both its ``main'' and ``deep'' surveys).
The dispersion in the Hubble diagram of the LSST SN catalog is not yet completely understood, but according to recent photometric surveys and rough estimations in the LSST white paper, it seems that a dispersion of 0.15 mag constant in redshift may be a reasonable hypothesis. Note that since we define $\sigma_{\rm int}$ to include noise, it corresponds to what is sometimes referred to as the total Hubble diagram dispersion (as opposed to the idealized intrinsic SN dispersion, not accounting for photometric redshift and other instrumental errors).

The other forecast used in this paper is WFIRST, a spectroscopic survey from NASA and its spectroscopy in the near infrared can potentially reduce the intrinsic dispersion in the Hubble diagram to around 0.11 mag (which we will again assume constant in redshift) and achieve very deep redshifts as shown in the distribution of 2700 supernovae plotted in Figure~\ref{fig:SNz}. A similar proposal to WFIRST is DESIRE \citep{Astier:2014swa}, which has a good overlap in redshift with WFIRST and seems to naturally complement LSST at higher redshifts. Here, however, we will  consider only LSST and WFIRST as they are good representatives of future surveys.

The JLA catalog, on the other hand, is a joint analysis of the 740 spectroscopically confirmed supernovae type Ia of the SNLS and SDSS-II collaboration. The redshift depth reaches the value of $1.3$, but for redshifts higher than unity the number of supernovae is insufficient to carry out the MeMo analysis. Here we use the same technical choices of \citealt{Castro:2014oja}, which are bins of $0.1$ width in redshift in the range $0<z<0.9$, yielding a total of $706$ supernovae.

%%%%%%%%%%%%%%%%%%%%%%%%%%%%%%%%%%%%
\subsection{Chandra low-redshift cluster sample}
\label{prior}
%%%%%%%%%%%%%%%%%%%%%%%%%%%%%%%%%%%%

Chandra observations put tight constraints on $\omegam$ and $\sig$ \citep{Vikhlinin:2008ym}. Here we will focus on the low-redshift cluster sample which has a median redshift of $z\approx \bar z \equiv 0.05$.
The posterior on $\omegam$ and $\sig$ is given in Fig.~3 of \citealt{Vikhlinin:2008ym}\footnote{This figure includes also the high-redshift sample. However, the constraints are dominated by the low-redshift sample.} and can be approximated by the following binormal distribution:
\begin{align}
L_{\rm cl} &\,=\, L_{{\rm cl},0}\, \exp \left[  - \frac{1}{2} (\boldsymbol x - \boldsymbol x_{\rm bf})^t \Sigma_{\rm cl}^{-1} (\boldsymbol x - \boldsymbol x_{\rm bf}) \right] \,,  \nonumber \\
\boldsymbol x &\,=\, \{ \omegam, \sig \} \, ,
\label{eq:clprior}
\end{align}
where $\Sigma_{\rm cl}$ is the covariance matrix (inverse of the Fisher matrix) which is determined by the dispersions $\sigma_{\omegam}=0.05$ and $\sigma_{\sig}=0.08$ and the correlation $\rho=-0.985$. The best-fit values are $\boldsymbol x_{\rm bf}= \{0.258, 0.8  \}$.
Also, we do not put prior constraints on $\Omega_{m0}$ as the latter will be already well constrained by supernovae data.
Although this is an approximation to the posterior in \citealt{Vikhlinin:2008ym}, it suffices for our scope which is to show the complementarity of SN lensing (using the JLA catalog in this case) to other probes of matter perturbations.

As the latter constraints on $\omegam$ and $\sig$ are basically at present time ($\bar z \ll1$), they depend weakly on the value of $\gamma$.
To nevertheless account for such dependence we can proceed as follows.
The constraint of Eq.~(\ref{eq:clprior}) is obtained at $\bar z$, and then evolved to $z=0$ using linear theory for $\gamma=\gas$.
In order to expand this likelihood into the $\{ \omegam, \sig, \gamma \}$ parameter space we have to evolve it back to $\bar z$ and then evolve it forward again to $z=0$ using the growth function specific to the wanted $\gamma$. This simply means that for each slice of $\gamma =$const one has to deform Eq.~(\ref{eq:clprior}) according to:
\begin{equation}
\siga = \frac{G_{\gas}(\bar z)}{G_{\gamma}(\bar z)} \; \sig \,,
\end{equation}
where $G$ is given in Eq.~(\ref{g-function}) and also depends on the value of $\omegam$.

\subsection{Growth rate data}
\label{growth}

Growth rate data measure the quantity $d = f(z)\sigma_{8}(z)= f(z)\sigma_{8}G(z)$, which depends on the three parameters $\,\Omega_{m0},\,\gamma,$ and $\sigma_{8}$, as  seen in Eqs.~\eqref{gammapar}--\eqref{g-function}. One can then build the following likelihood function:
\begin{equation}
L_{\rm gr}\,=\,L_{{\rm gr}, 0}\,\exp\left[-\frac{1}{2}(d_{i}-t_{i})C_{ij}^{-1}(d_{j}-t_{j})\right] \,,
\label{chiquadro-1}
\end{equation}
where $C_{ij}$ is the covariance matrix of the data and $t_i$  the theoretical predictions. It is important to remark that also for the growth rate data  we  put practically no prior constraint on $\Omega_{m0}$ (i.e. uniform prior between 0.05 and 0.95), since $\Omega_{m0}$ will be already severely constrained by supernovae data.

We collected all the current independent published estimates of $f\sigma_{8}(z)$ obtained with the redshift space distortion method from 2dFGS~\citep{Percival:2004fs}, 6dFGS~\citep{Beutler:2012px}, LRG \citep{Samushia:2011cs,Chuang:2012qt}, BOSS \citep{Tojeiro:2012rp}, CMASS~\citep{Samushia:2013yga,Chuang:2013wga}, WiggleZ \citep{Blake:2012pj} and VIPERS \citep{delaTorre:2013rpa} (see
also~\cite{Macaulay:2013swa,Beutler:2013yhm,Reid:2014iaa,More:2014uva}).  All together they cover the redshift interval $[0.07,0.8]$. In some cases the correlation coefficient between two samples has been estimated in~\citealt{Macaulay:2013swa} and included in our analysis. In the following we will refer to this data with ``RSD''.

It is important to note that the RSD data are in principle partially degenerated with the Alcock-Paczyński (AP) effect~\citep{Alcock:1979mp}, which is the fact that spherical objects do not appear spherical to observers if the wrong cosmology is assumed. One way to take the AP effect into account is to marginalize over the AP parameters, which generally enlarges the error bars but remove possible biases. All the data we use (listed above) do this either explicitly or implicitly.

We also estimated the accuracy of the estimation of $f\sigma_{8}(z)$ obtained from redshift distortions  in the redshift range $[0.5,2.1]$ in  a future survey that approximates the Euclid mission \citep{Euclid-r,Amendola:2012ys}; this
has been obtained in \citealt{Amendola:2013qna}, to which we refer for all the exact specifications. In the following we will refer to this data with ``Euclid''. Needless to say, the Euclid mission will provide much more cosmological information, but here we will utilize only RSD data because in a $\Lambda$CDM background they depend only on $\Omega_{m0},\,\gamma$ and $\sigma_{8}$.

To sum up, we will focus our study on the following data combinations:

\begin{itemize}

\item \emph{current}: the JLA supernova catalog with the Chandra low-redshift cluster sample (see Fig.~\ref{fig:SN+clus}) and the current RSD data (see Fig.~\ref{fig:SN+grow}),

\item \emph{future}: the LSST (and also WFIRST) supernova catalog with Euclid-like forecast data (see Fig.~\ref{fig:SN+grow-future}).

\end{itemize}

%%%%%%%%%%%%%%%%%%%%%%%%%%%%%%%%%%%%
%%%%%%%%%%%%%%%%%%%%%%%%%%%%%%%%%%%%
\section{Constraints on \texorpdfstring{\MakeLowercase{$f_g$}}{$f_g$} and $\sig$}
\label{constraints}
%%%%%%%%%%%%%%%%%%%%%%%%%%%%%%%%%%%%
%%%%%%%%%%%%%%%%%%%%%%%%%%%%%%%%%%%%

For simplicity and numerical convenience we will fix all the cosmological parameters to the Planck 2013 best-fit values (see Section~\ref{grope}). Only $\sig$, $\omegam$ and the growth-rate parameters will be let free. This is justified by the fact that lensing depends weakly on parameters other than these \citep{Amendola:2013twa}. In fact, $\omegam$ is already presently constrained at the $12\%$~level by SN data \citep{Betoule:2014frx} and will be much more so by future LSST supernova data  (the exact precision cannot be easily forecast as it will likely be systematics dominated) and by Euclid. Thus, whenever we use LSST data it is in practice irrelevant whether we marginalize over or fix the tightly constrained $\omegam$.
Moreover, as shown by \citealt{2010MNRAS.406.1796R}, $\omegam$ is not strongly correlated with $\gamma$. In other words, the $\gamma$-parametrization allows to probe departures from GR, independently from the modeling of the background ($\omegam$ in this case).
Nevertheless, in order to obtain conservative results, when using JLA data we always marginalize over $\omegam$.

%%%%%%%%%%%%%%%%%%%%%%%%%%%%%%%%%%%%
%%%%%%%%%%%%%%%%%%%%%%%%%%%%%%%%%%%%
\subsection{Current constraints}
\label{c-constraints}
%%%%%%%%%%%%%%%%%%%%%%%%%%%%%%%%%%%%
%%%%%%%%%%%%%%%%%%%%%%%%%%%%%%%%%%%%

The non-Gaussian lensing scatter in present SN catalogs cannot yet put significant bounds in perturbation quantities, as much more statistics is needed. Nevertheless, it is interesting to investigate the current confidence levels for at least two reasons.
The first is that it illustrates the improvements future probes can bring to this analysis, if systematics can be kept under control.
The second is that it allows to test whether systematics are already biasing the current results.

\begin{figure}
\begin{centering}
\includegraphics[width=.9\columnwidth]{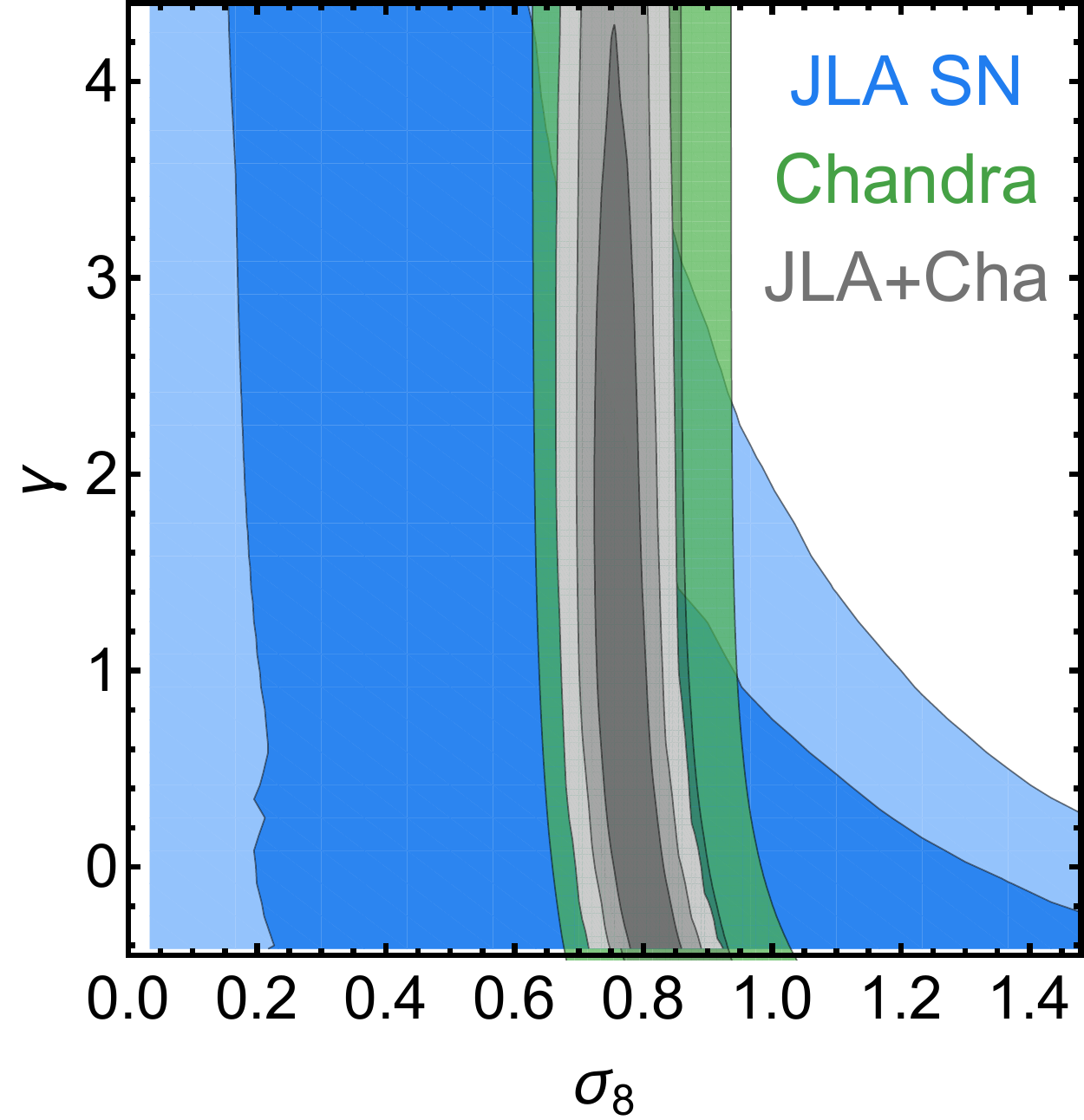}
\caption{1 and 2$\sigma$ marginalized constraints on $\gamma$ and $\sig$ for the JLA supernova catalog and the Chandra low-$z$ cluster data. The combined contours are only capable of constraining $\sigma_8$, and just add very loose bounds on $\gamma$. See Section~\ref{c-constraints} for more details and Table~\ref{c-tab} for the marginalized constraints.}
\label{fig:SN+clus}
\end{centering}
\end{figure}

\begin{figure}
\begin{centering}
\includegraphics[width=.9\columnwidth]{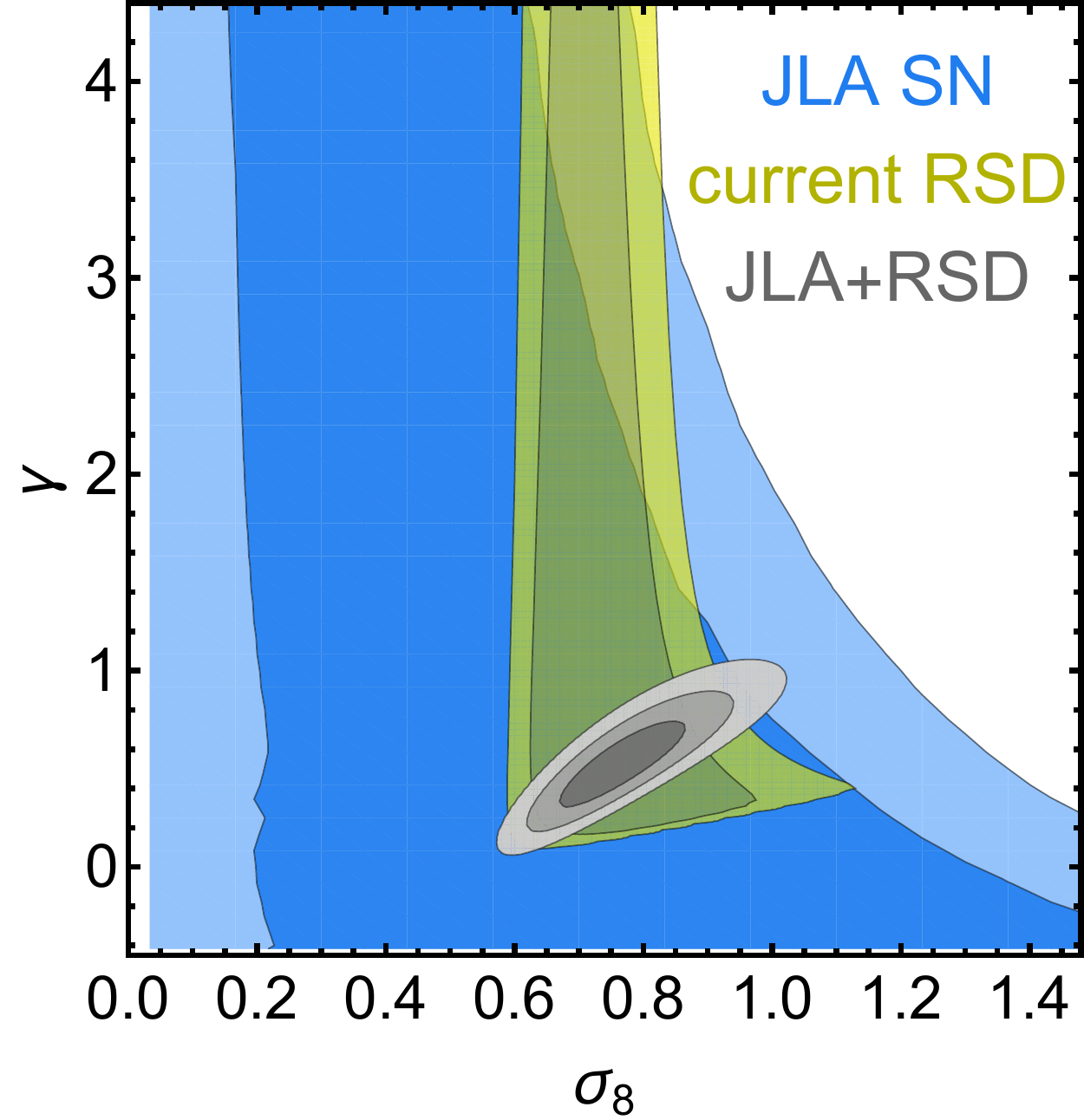}
\caption{Similar to Figure~\ref{fig:SN+clus} but now combining JLA SNe with present redshift space distortion (RSD) data (see Section~\ref{growth}). The combined contours are much smaller than what could be inferred by eye because the growth rate constraints depend strongly on the prior on $\omegam$, which present SNe already constrain quite well. As explained in the text we adopt here a very broad $0.05<\omegam<0.95$ prior. The same remark holds for Fig.~\ref{fig:SN+grow-future}.}
\label{fig:SN+grow}
\end{centering}
\end{figure}

Figs.~\ref{fig:SN+clus} and~\ref{fig:SN+grow} show the constraints on $\sig$ and $\gamma$ for the combinations of JLA data with the Chandra low-redshift cluster sample and RSD growth rate data, respectively. The SN constraints have been marginalized over the second and third intrinsic moments, as discussed at the end of Section~\ref{semo}.
For the growth data we assumed a flat prior corresponding to $0.05 < \omegam < 0.95$, which effectively corresponds to $0.05 < \omegam < \infty$ (i.e., basically we only assume the baryons density has been constrained by e.g.~big-bang nucleosynthesis measurements). As it can be seen, current SN data seems to be perfectly unaffected by systematics, and can already help improve the constraints obtained by either of the other two techniques alone. It is important to note, however, that the improvement in the contours comes mostly from the fact that SN constrains $\omegam$ much better than the other techniques, while lensing currently plays only a minimal constraining role.

Marginalized constraints can be found in Table~\ref{c-tab}.

%%%%%%%%%%%%%%%%%%%%%%%%%%%%%%%%%%%%
%%%%%%%%%%%%%%%%%%%%%%%%%%%%%%%%%%%%
\subsection{Future constraints}
\label{f-constraints}
%%%%%%%%%%%%%%%%%%%%%%%%%%%%%%%%%%%%
%%%%%%%%%%%%%%%%%%%%%%%%%%%%%%%%%%%%

%%%%%%%%%%%%%%%%%%%%%%%%%%%%%%%%%%%%
\subsubsection{$\gamma$-parametrization}
\label{gammacs}
%%%%%%%%%%%%%%%%%%%%%%%%%%%%%%%%%%%%

\begin{figure}
\begin{centering}
\includegraphics[width=0.95\columnwidth]{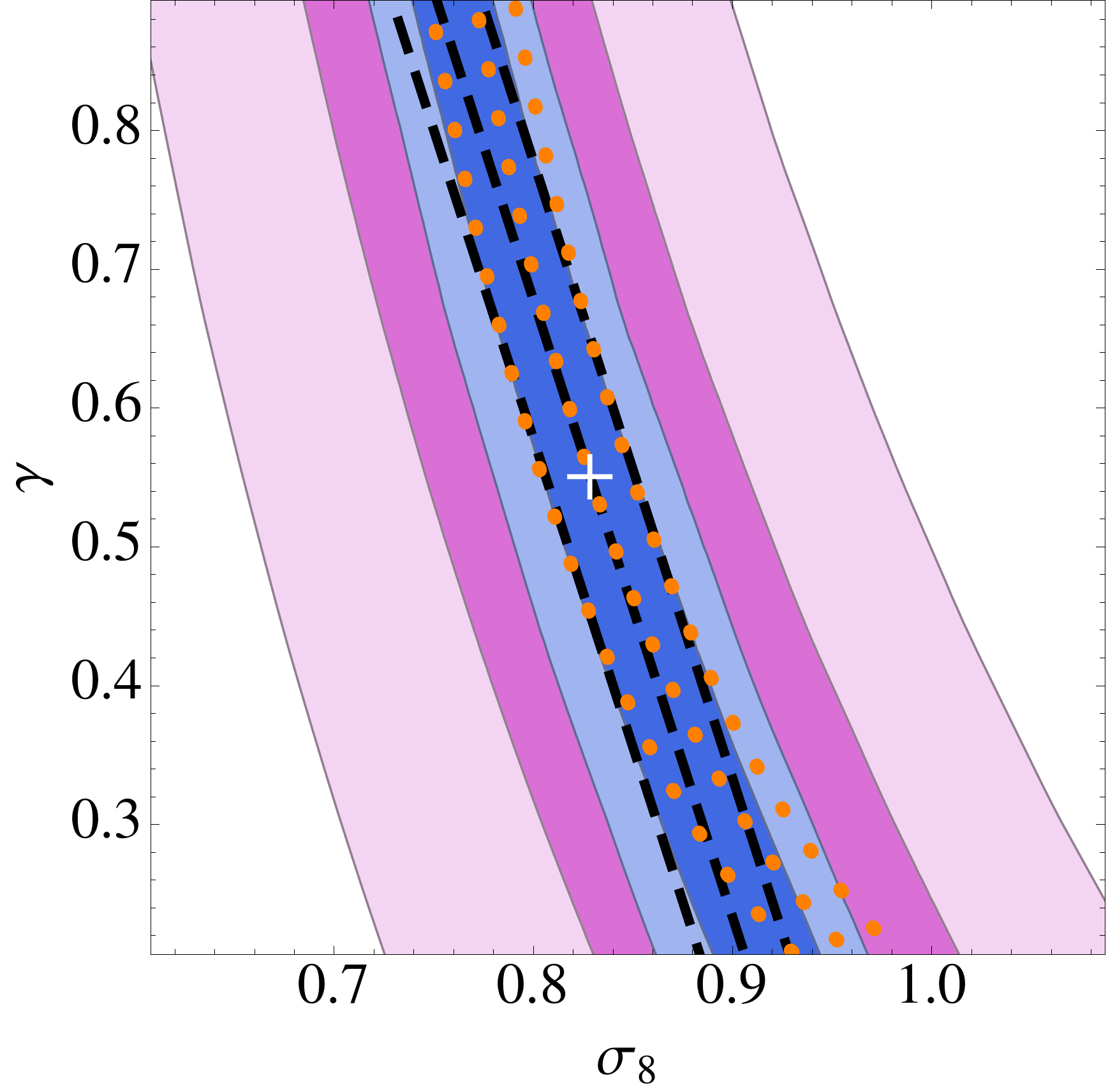}
    \caption{1 and 2$\sigma$ marginalized constraints on $\gamma$ and $\sig$ using the LSST (blue contours) and WFIRST (orchid contours) supernova catalogs of Fig~\ref{fig:SNz}. Empty contours show the parametrizations for the degenerate constraints on $\gamma$ and $\sig$ given by Eq.~(\ref{par1}) (orange dotted lines) and Eq.~(\ref{par2}) (black dashed lines). The fiducial model $\{\sigma_{8, \rm{fid}} , \gamma_{\rm{fid}}\} =\{0.83, 0.55 \}$ is marked with a white cross.}
\label{fig:degline}
\end{centering}
\end{figure}

In Fig.~\ref{fig:degline} constraints on $\gamma$ and $\sig$ using  synthetic catalogs from LSST and WFIRST are shown.
These posteriors have been marginalized over the second-to-fourth intrinsic moments, but not over $\omegam$ as LSST  will tightly constrain it and results are unchanged if $\omegam$ is simply kept fixed.
As commented at the beginning of Section~\ref{constraints}, $\omegam$ is also not strongly correlated with $\gamma$ and thus these constraints are expected to be valid even in the case that the constraints from LSST on $\omegam$ get compromised because of systematics. WFIRST constraints have proven not to be competitive. In the remaining of this paper we will only use  the LSST catalog to make forecasts.

The constraints of Fig.~\ref{fig:degline} can be summarized using either of the following two parametrizations:
\begin{align}
\gamma   \, \left (\frac{\sig}{\sigma_{8, \rm{fid}}} \right)^\alpha &\,=\, \gamma_{\rm{fid}} \pm \sigma_\gamma \,,  \label{par1} \\
\gamma  -\gamma_{\rm{fid}} \left (1- \frac{\sig}{\sigma_{8, \rm{fid}}} \right) \alpha &\,=\, \gamma_{\rm{fid}} \pm \sigma_\gamma \,, \label{par2}
\end{align}
where, in both cases, $\alpha \approx 6.7$ and $\sigma_\gamma \approx 0.1$.
These parameterizations are valid near the (best fit) fiducial model and are shown as empty contours in Fig.~\ref{fig:degline}. Parametrization \eqref{par2} performs better than \eqref{par1} for very low values of $\gamma$. The result of Eq.~(\ref{par1}) can be compared with the results of \citealt{2010MNRAS.406.1796R}, where the combination of XLF, $f_{\rm gas}$, SN Ia, BAO and CMB data has given the constraint $\gamma (\sig / 0.8)^{6.8} =0.55^{+0.13}_{-0.10}$. This shows how future LSST constraints alone could be as competitive as all the present-day constraints (considered in \citealt{2010MNRAS.406.1796R}) combined.

\begin{figure}
\begin{centering}
\includegraphics[width=.95\columnwidth]{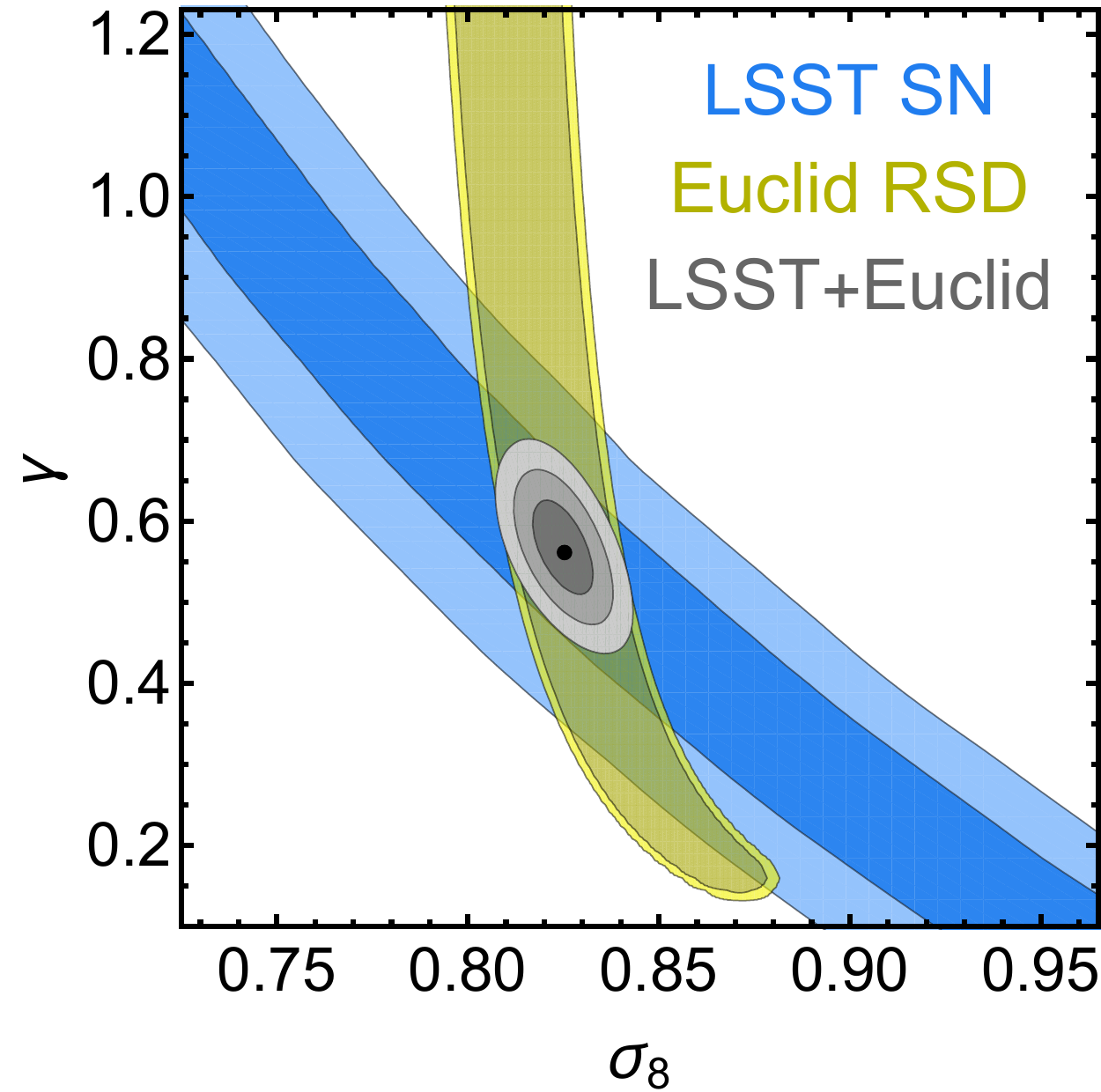}
\caption{Similar to Figure~\ref{fig:SN+grow}, this time forecasting future SN and growth of structure data. For the SNe we use the LSST catalog assuming 500,000 SNe, each with a total Hubble diagram scatter of 0.15 mag; for the growth of structure, we use  forecast data for a Euclid-like probe with a $0.05<\Omega_{m0}<0.95$ prior.
See Section~\ref{gammacs} for more details and Table~\ref{c-tab} for the marginalized constraints.}
\label{fig:SN+grow-future}
\end{centering}
\end{figure}

\begin{table}
\begin{center}
    \begin{tabular}{ccccccc}
    \hline
    \hline
    & \phantom{b} & JLA+Chandra & \phantom{b} & JLA+RSD & \phantom{b}& LSST+Euclid \\
\hline
   $\sig$ & & $0.77^{+0.03}_{-0.04}$ & & $0.76^{+0.07}_{-0.06}$ & & $0.83\pm0.005$ \\
 $\gamma$ & &  unconstrained & &  $0.52^{+0.16}_{-0.13}$ & &  $0.55\pm0.04$ \\
    \hline
    \hline
    \end{tabular}
\end{center}
\vspace{-.2cm}
\caption{One-dimensional 1$\sigma$ constraints (marginalized over the remaining parameters) relative to Figs.~\ref{fig:SN+clus}, \ref{fig:SN+grow} and \ref{fig:SN+grow-future}.\vspace{.1cm}\label{c-tab}}
\end{table}%

Fig.~\ref{fig:SN+grow-future}  shows LSST and Euclid-like  constraints on $\sig$ and $\gamma$, and also the joint contours. The Euclid-like constraints are shown for the case in which the posterior is marginalized over $\omegam$ with a flat prior $0.05 < \omegam < 0.95$ (although here, RSD data does allow higher $\omegam$). Marginalized constraints can be found in Table~\ref{c-tab}. Although once again the majority of the constraining power of SN comes from the nailing down of $\omegam$, the SN lensing contours (from the higher moments) start to become competitive, and pose an important cross-check to the more standard techniques.

In fact, this plot clearly shows how -- in the quest for stricter dark energy constraints -- one should rely on different probes as they suffer different parameter degeneracies and can efficiently complement each other. Also, different probes are subject to different systematic uncertainties, and any new probe is obviously welcome in order to cross-check the validity of other measurements. The importance of such complementarity regarding constraints on $\sig$ and $\gamma$ has been exemplified by \citealt{2013MNRAS.432..973R} where a combination of galaxy growth, CMB and cluster growth observables was able to completely break the degeneracy.

%\clearpage
%%%%%%%%%%%%%%%%%%%%%%%%%%%%%%%%%%%%
\subsubsection{$f$-parametrization}
\label{fpares}
%%%%%%%%%%%%%%%%%%%%%%%%%%%%%%%%%%%%

\begin{figure}
\begin{centering}
    \includegraphics[width=.95\columnwidth]{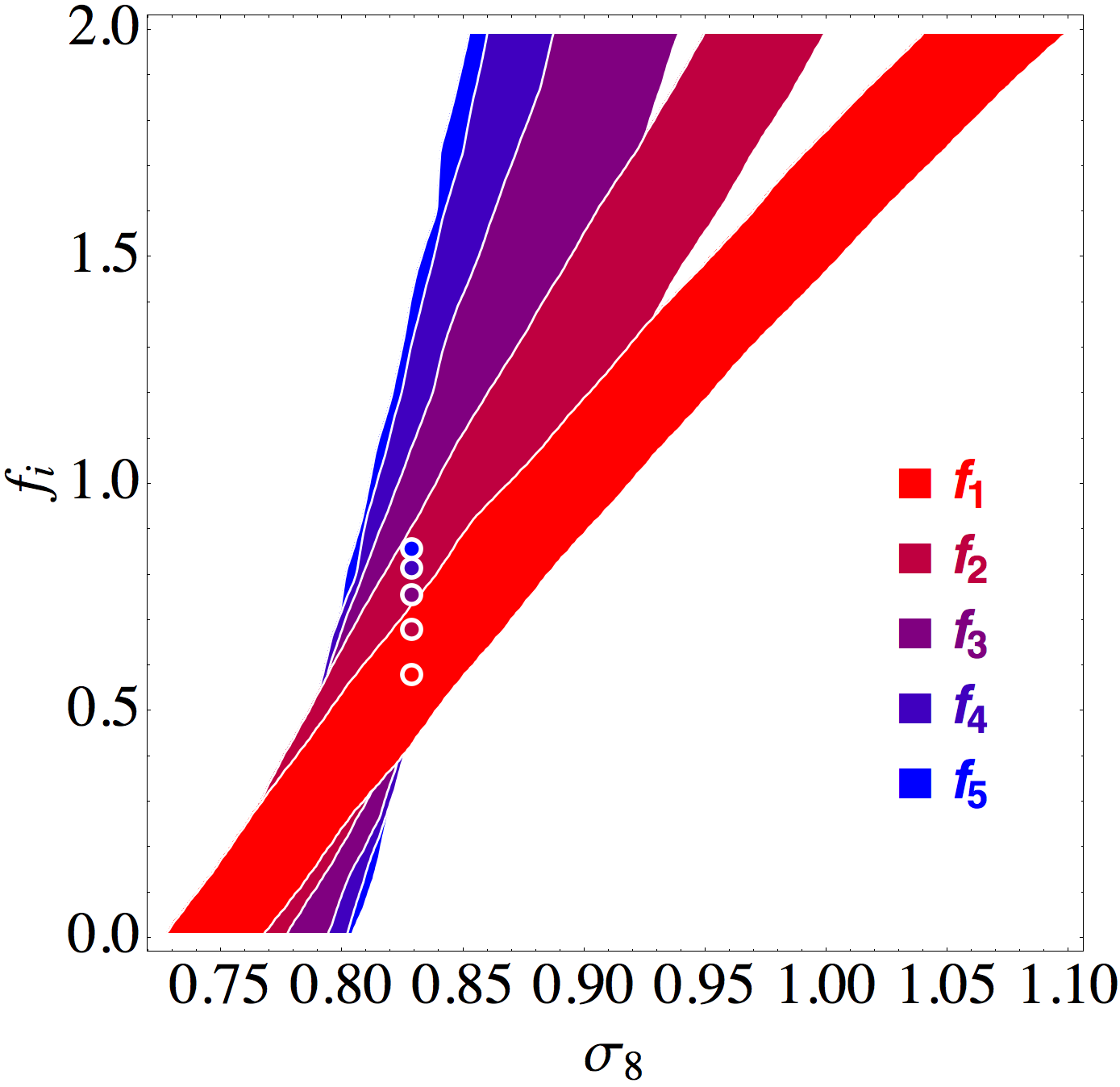}
    \caption{1$\sigma$ marginalized constraints on $\sig$ and $f_i$ of Eq.~(\ref{fpareq}) using the LSST catalog. The binned values of the growth rate are varied one at time, the others are assumed to take the standard values listed in Table~\ref{fbins}. See Section~\ref{fpares}.% for more details.
    }
\label{fig:fcombo}
\end{centering}
\end{figure}

Here we forecast constraints on the binned values $f_{\:\!\bar i}$ of Eq.~(\ref{fpareq}), taken one at time. Bins for which $i \neq \bar i$ are assumed to take the standard values listed in Table~\ref{fbins}.
Fig.~\ref{fig:fcombo} shows how the 68\% constraints (marginalized over intrinsic moments) rotate and get weaker as the corresponding redshift increases. Constraints get weaker because lensing is an integrated effect and will differ more with respect to the standard model if the change in the growth rate extends for a larger redshift range (see e.g.~Fig.~\ref{fig:fG}). Constraints rotate because lensing is less sensitive with respect to changes in growth rate at larger redshifts.
Also, it is interesting to note that -- if let free -- $f_{\:\!\bar i}$ and $f_{\:\!\bar j}$ are anti-correlated. This is expected as an increase in e.g.~$f_{\:\!\bar i}$ needs to be compensated by a decrease in $f_{\:\!\bar j}$ in order to have an equivalent growth of structure.
As mentioned before, this analysis could have been done using a different gravity theory, and Fig.~\ref{fig:fcombo} also shows that supernova lensing analysis can potentially be another way to test modified gravity, being more sensible to models that predict deviation from the standard model at smaller redshifts.

These LSST constraints can be summarized using the following parametrization:
\begin{align}
    f_i  -f_{i,\rm{fid}} \left (1- \frac{\sig}{\sigma_{8, \rm{fid}}} \right) \alpha_i &= f_{i,\rm{fid}} \pm \sigma_i \,, \label{pardegf}
\end{align}
where the value of $f_{i,\rm{fid}}$ are given in Table~\ref{fbins} and the parameters $\alpha_i$ and $\sigma_i$ by the following linear fit:
\begin{align}
\alpha_i  &= -8.9 - 12 \, \bar z_i   \,, \\
\sigma_i &= 0.12 + 0.35 \, \bar z_i \,,
\end{align}
where $\bar z_i$ is the redshift value at the center of the redshift bin $[z_{i-1}, z_i)$. These constraints should be valid for bins at redshifts $z \neq \bar z_i$ as long as the bin has a width of $\Delta z=0.2$.
As in Section \ref{gammacs}, these constraints have been marginalized over the second-to-fourth intrinsic moments but not over $\omegam$, again because for LSST this is irrelevant.

%%%%%%%%%%%%%%%%%%%%%%%%%%%%%%%%%%%%
%%%%%%%%%%%%%%%%%%%%%%%%%%%%%%%%%%%%
\section{Conclusions}
\label{conclusions}
 %%%%%%%%%%%%%%%%%%%%%%%%%%%%%%%%%%%%
%%%%%%%%%%%%%%%%%%%%%%%%%%%%%%%%%%%%

In this paper we studied the current and future constraints on the linear matter growth rate and on the power spectrum normalisation $\sigma_8$ using supernova lensing through a recently proposed technique combined with other data. The growth rate is parametrized either in the popular form $f=\Omega_m^\gamma$ or as a stepwise constant function. For the present data, we included datasets taken from the JLA supernova Ia catalog, from the Chandra low-redshift cluster sample, and from all the available $f\sigma_8 (z)$ redshift-distortion data. The final results, summarized in Table~\ref{c-tab}, show that $\sigma_8$ can be constrained by the combined datasets (in particular by JLA and RSD) to within 10$\%$ roughly, while $\gamma$ is only poorly constrained to within 30$\%$ (all errors at 1$\sigma$). Current supernova data alone put only very broad constraints on $\gamma,\sigma_8$ and the results are driven by the other datasets.

For the future constraints we show that, with 500,000 LSST supernovae with average total dispersion of $0.15$ mag, the constraints on $\gamma,\sigma_8$ will lie on a band parametrized by Eq. (\ref{par1}) or  (\ref{par2}), see Fig. \ref{fig:SN+grow-future}. It is worth stressing that future LSST constraints alone can potentially be as competitive as all present-day constraints together. When combined with  Euclid-like results on $f\sigma_8 (z)$ from redshift distortions, the parameters will be constrained to within $0.6\%$ and 7$\%$ on $\sigma_8$ and $\gamma$, respectively  (see Table~\ref{c-tab}). We also explored the constraints on a general, step-wise parametrization of the growth rate $f$.

The  three Figures~\ref{fig:SN+clus},~\ref{fig:SN+grow} and \ref{fig:SN+grow-future} nicely show the twofold beneficial effect of SN lensing analysis: at the background level (the first moment of the lensing PDF) it breaks the degeneracy on $\omegam$ and $\sig$ -- this is why we obtain better results than what would naively be inferred by eye -- and at the perturbation level (the higher moments) it breaks the degeneracy on $\gamma$ and $\sig$.

The main interest in this method based on supernova lensing is that it exploits an effect completely different from the standard ones based on clusters abundances,  galaxy clustering,  weak lensing and  strong lensing abundances, and therefore subject to different systematics. An additional bonus of our method is that  a deviation from the parameters that fit other probes could signal for instance a redshift dependence of the supernovae magnitude  central moments, which could then be related to redshift-dependent physics of supernova lightcurves.

\section*{Acknowledgments}
It is a pleasure to thank Chia-Hsun Chuang and David Rapetti for fruitful discussions.
LA acknowledges support from DFG through the TRR33 program ``The Dark Universe''.
TC is grateful to Brazilian research agency CAPES. VM was supported during part of this project by a Science Without Borders fellowship from the Brazilian Foundation for the Coordination of Improvement of Higher Education Personnel (CAPES). MQ is grateful to Brazilian research agencies CNPq and FAPERJ for support.

\appendix

\section{Fitting functions}
\label{app:fits}

Here we will give simple analytical fitting functions for the second-to-fourth central lensing moments $\mu_{2-4, \text{lens}}$ as a function of
$\{z,\,\sigma_{8},\,\gamma\}$, which are valid within the domain:
\begin{align*}
    0 \le  & \;\,z\,  \le  1.2 \,,   \\
    0.6 \le & \;\sigma_{8} \!\le   1  \,, \\
    0  \le & \;\,\gamma\, \le  1.1 \,.
\end{align*}
All the other cosmological parameters have been fixed to the Planck 2013 best-fit values (see Section~\ref{grope}).
Using magnitudes, the fitting formulae are:
\begin{align}
    \sigma_{\rm lens}(z, \sigma_8, \gamma) &=\sigma_8 z (0.0164 \gamma +0.0145 z+0.0396) \, ,  \label{2fit}\\
    \mu_{3,{\rm lens}}^{1/3}(z, \sigma_8, \gamma) &=\sigma_8 z (0.0283 \gamma -0.00696 z+0.0861) \,, \label{3fit} \\
    \mu_{4,{\rm lens}}^{1/4}(z, \sigma_8, \gamma) &=\sigma_8 z (0.0393 \gamma -0.0344 z+0.153) \,. \label{4fit}
\end{align}
In the entire domain of validity, the average RMS error is 0.0012, 0.0017 and 0.0021 for $\mu_{2-4, \text{lens}}$, respectively, which is roughly 3\% for all three moments.

%%%%%%%%%%%%%%%%%%%%%%%%%%%%%%%%%%%%
%%%%%%%%%%%%%%%%%%%%%%%%%%%%%%%%%%%%
\bibliographystyle{mn2e_eprint}
\bibliography{cosmo-lensing,growth-rate}
 %%%%%%%%%%%%%%%%%%%%%%%%%%%%%%%%%%%%
%%%%%%%%%%%%%%%%%%%%%%%%%%%%%%%%%%%%

%%%%%%%%%%%%%%%%%%%%%%%%%%%%%%%%%%%%
%%%%%%%%%%%%%%%%%%%%%%%%%%%%%%%%%%%%

\label{lastpage}
\bsp
\end{document}